\documentclass[letterpaper]{article} 
\usepackage{aaai2026}  
\usepackage{times}  
\usepackage{helvet}  
\usepackage{courier}  
\usepackage[hyphens]{url}  
\usepackage{graphicx} 
\usepackage{natbib}  
\usepackage[utf8]{inputenc}   

\usepackage{caption} 
\usepackage{algorithm}
\usepackage{algorithmic}

\usepackage{newfloat}
\usepackage{listings}
\DeclareCaptionStyle{ruled}{labelfont=normalfont,labelsep=colon,strut=off} 
\floatstyle{ruled}
\newfloat{listing}{tb}{lst}{}
\floatname{listing}{Listing}
\nocopyright

\title{When AI Trading Agents Compete: Adverse Selection of Meta-Orders by Reinforcement Learning-Based Market Making}

\author{
    Ali Raza Jafree\textsuperscript{\rm 1},
    Konark Jain\textsuperscript{\rm 1, 2}\thanks{Opinions expressed in this paper are those of the authors, and do not necessarily reflect the view of JP Morgan.},
    Nick Firoozye\textsuperscript{\rm 1}
}
\affiliations{
    \textsuperscript{\rm 1}Department of Computer Science, University College London, London, UK\\
    \textsuperscript{\rm 2}Quantitative Research, JP Morgan Chase, London, UK\\

    konark.jain.23@ucl.ac.uk

}

\usepackage{bibentry}

\begin{document}

\maketitle

\begin{abstract}
We investigate the mechanisms by which medium-frequency trading agents are adversely selected by opportunistic high-frequency traders. We use reinforcement learning (RL) within a Hawkes Limit Order Book (LOB) model in order to replicate the behaviours of high-frequency market makers. In contrast to the classical models with exogenous price impact assumptions, the Hawkes model accounts for endogenous price impact and other key properties of the market \cite{Jain2024}. Given the real-world impracticalities of the market maker updating strategies for every event in the LOB, we formulate the high-frequency market making agent via an impulse control reinforcement learning framework \cite{jain2025impulse}. The RL used in the simulation utilises Proximal Policy Optimisation (PPO) and self-imitation learning. To replicate the adverse selection phenomenon, we test the RL agent trading against a medium frequency trader (MFT) executing a meta-order and demonstrate that, with training against the MFT meta-order execution agent, the RL market making agent learns to capitalise on the price drift induced by the meta-order. Recent empirical studies have shown that medium-frequency traders are increasingly subject to adverse selection by high-frequency trading agents. As high-frequency trading continues to proliferate across financial markets, the slippage costs incurred by medium-frequency traders are likely to increase over time. However, we do not observe that increased profits for the market making RL agent necessarily cause significantly increased slippages for the MFT agent. 
\end{abstract}

\begin{links}
    \link{Code}{https://github.com/konqr/lobSimulations/tree/master/HawkesRLTrading}
\end{links}

\section{Introduction}
In this study, we use the LOB simulation by Jain and Firoozye \cite{Jain2024, Jain2024Review, jain2024no} to construct a simulated trading environment in which a medium-frequency trader faces strategic adversarial selection from HFTs, replicating the empirically observed phenomenon of medium-frequency traders (MFTs) being adversely selected by high-frequency traders (HFTs) \cite{carteaCollusion}. The LOB is simulated using a compound Hawkes process, which differs from traditional models which are generally based on Brownian motion; most significantly, it accounts for the market impact of endogenous factors \cite{jain2025impulse}. That is, the simulation responds to orders and trades made by the trading agents, making it suitable for our goal to effectively replicate the behaviour of HFTs adversely selecting MFTs. 

Our methodology encompasses three phases. Firstly, we develop trading agents within the LOB to simulate MFT meta-order flows and a HFT market maker. Additionally, testing is required to establish a baseline for the MFT's slippages and the HFT's performance. For the MFT agent, we chose to create an agent that executes a meta-order using the time-weighted average price (TWAP) execution method. While TWAP is not the most sophisticated meta-order execution method, it is still actively used in less developed markets and as a part of quantitative trading strategies \cite{algotradingquantstrats}. To create a trading agent that accurately simulates a HFT market maker, we utilise the RL trading agent framework previously developed \cite{jain2025impulse}. The RL framework approximates the solution to the high dimensional Hamilton-Jacobian-Bellman Quasi-Variational Inequality (HJB-QVI), brought about by the fact that a market maker will only respond to a subset of all LOB events due to real-world technological limiting factors and practical constraints. The RL model's enabling of the agent to have a higher degree of decision making when it comes to when orders are placed further enhances its accuracy, modelling the precise and purposeful intervention decisions made by real-world market makers.

Secondly, to test the agents when trading together and investigate whether the 'unaware' RL (uRL) trading agent could, by default, capitalise on the TWAP meta-order execution without additional training. 

Finally, to train an RL model with the meta-order execution agent, giving it full information as to whether the TWAP is present or not and whether it is executing a buy meta-order or a sell meta-order. 

The eventual goal of this work is to develop defensive mechanisms for medium-frequency traders and their meta-order executions against adverse selection by the RL trained and enabled HFT.

\section{Background}
\label{sec:background}

For readers unfamiliar with high-frequency trading and market microstructure, we briefly introduce the key concepts underlying this work.

\subsection{Limit Order Book (LOB)}
A limit order book is the core mechanism of modern electronic markets. It contains two lists of pending orders: buy orders (bids) and sell orders (asks). The highest bid and lowest ask define the best bid and ask, and their difference is the bid-ask spread.

Traders interact with the LOB through two main order types. A {limit order (LO)} is passive—it rests in the book at a chosen price until matched. A {market order (MO)} is aggressive—it executes immediately against the best available prices. The flow of order arrivals, cancellations, and executions determines both price movements and liquidity.

\subsection{Market Making}
Market makers continuously post buy and sell limit orders, earning profits by capturing the bid-ask spread—buying at the bid and selling at the ask. They face several risks: {inventory risk} (holding large unbalanced positions), {adverse selection risk} (trading against informed agents), and {execution risk} (orders not being filled).

High-frequency market makers operate on millisecond timescales, constantly adjusting quotes in response to market changes using automated algorithms and low-latency infrastructure.

\subsection{Meta-orders and Execution Strategies}
Large institutional investors must often trade amounts far exceeding available liquidity. Executing such “meta-orders” at once would cause strong price impact—buying pushes prices up, selling pushes them down.

To limit impact, these orders are split into smaller “child” orders executed over time via algorithms such as:
\begin{itemize}
\item \textbf{TWAP:} Time Weighted Average Price execution - executes equal-sized slices at regular intervals.
\item \textbf{VWAP:} Volume Weighted Average Price execution - follows historical volume patterns.
\item \textbf{POV:} Percentage of Volume based execution - maintains a constant share of total market volume.
\end{itemize}
While these methods reduce costs, their predictable patterns can be detected and exploited by faster traders.

\subsection{Adverse Selection}
Adverse selection arises when an agent trades with better-informed agents, leading to losses. During meta-order execution, high-frequency traders may detect and anticipate the large order’s direction, profiting from the predictable price drift.

For the slower trader, this results in {slippage}—the difference between the expected and actual execution price. As markets become faster and more competitive, such adverse selection effects intensify.

\subsection{Hawkes Processes}
Simple Poisson or Brownian models fail to capture the clustering and feedback seen in real order flows. Hawkes processes, by contrast, are self-exciting point processes where each event increases the likelihood of future ones.

In the LOB context, this means market and limit order arrivals trigger further activity, producing realistic volatility clustering and feedback. Our compound Hawkes process model captures key empirical features:
\begin{itemize}
\item \textbf{Endogenous price impact:} Price changes emerge from order flow dynamics.
\item \textbf{Self-excitation:} Trading activity breeds more trading activity.
\item \textbf{Mean reversion:} Temporary impact decays over time.
\end{itemize}
This modelling framework allows us to study how strategic trading actions and endogenous feedback jointly shape market dynamics. Research has also shown that foundational models can be used with retrieval augmented generation, as shown in \cite{orgGlaze_MMRAugGen}.

\section{Related Work}

Our work primarily relates to the pool of research on interactions between HFTs and non-HFTs. Most recently, \cite{carteaCollusion} found that HFTs collude within the limit order book and adversely select non-HFTs. Whereas \cite{yao_and_ye} observed that an elevated risk of trading with informed participants correlates with wider bid-ask spreads, enabling non-HFTs to potentially secure execution priority through price mechanisms, research seems to show that HFTs can enhance overall market liquidity and quality. \cite{brogaard_hft_pricediscov} found that the contribution of HFTs towards price discovery is more significant than that of non-HFTs, specifically within the realm of limit orders. Similarly, \cite{baldauf_mollner_HFT} demonstrated HFTs' proficiency in order anticipation while concurrently improving market liquidity by narrowing bid-ask spreads. However, other research such as \cite{zhang_HFT} and \cite{caivano_HFT_volatility} suggests they may simultaneously introduce heightened market volatility. 

The expanding body of literature regarding the market impact of metaorders and high-frequency trading is also highly relevant to this research. \cite{donier2015fullyconsistentminimalmodel} found that, by using a diffusion model and neglecting fluctuations, the price impact of a metaorder can be computed exactly. Alternatively, \cite{naviglio2025estimationmetaorderimpactpublic} proposed a linear model to provide price trajectories of metaorders, with other models being shown to overestimate the positive triggering effect that a child order has on the order flow of the market. Recently, \cite{maitrier2025generatingrealisticmetaorderspublic} used public trade data to generate viable synthetic metaorders which exhibit all stylised facts of metaorder impact. 

The field of RL-driven algorithmic trading strategies is also pertinent to this work. \cite{Colliard2022-rd} studied the market impact of RL-driven MM algorithms, and found that the algorithms learn to quote supra-competitive prices. \cite{gasperov_market_2021} showed how RL can be used in addition to an adversarial agent in order to make a trading algorithm robust to attacks.

\section{Trading agents}

In the Hawkes limit order book simulation, trading agents are programmed with an action frequency $f$, and are called to take an action every $f$ seconds. In addition to this, agents may be called whenever a market order (MO) is made or whenever the bid-ask spread changes. 

There are 12 action options in total that an agent can take in the LOB simulation. The agent can place bid or ask limit orders (LOs) or MOs; the LO can be deep, top-of-book, or in-spread. The action can also be to cancel a previously placed limit order at either level. The agent also has the option to not take an action; to skip its turn. In the following, the algorithms for the two agents in the simulator and their base performances are described. 


\subsection{TWAP Agent}


TWAP is a simple meta-order execution method. The agent executes a metaorder by executing child orders at fixed time intervals over a predetermined length of time. The TWAP agent's simplicity makes for predictable behaviour. For example, the average size of a child order $q$ can be trivially computed, for $Q$ the total size of the meta-order and $T$ the total time the agent has to execute: $E[q] = \frac{Q}{Tf}$. where $f$ is the agent's action frequency. The algorithm is developed in such a way that the actual sizes of the child orders should not deviate significantly from this value. 

Because placing a LO does not guarantee execution, the TWAP agent has to place MOs in addition to limit orders to execute the metaorder in the required total time. The total execution time and quantity is split up into smaller windows, with the agent placing market orders near the end of the window, depending on how many limit orders have been filled in order to ensure that the entire window volume is executed in the window time. 

\begin{figure}[h]
    \centering
    \includegraphics[width=1\linewidth]{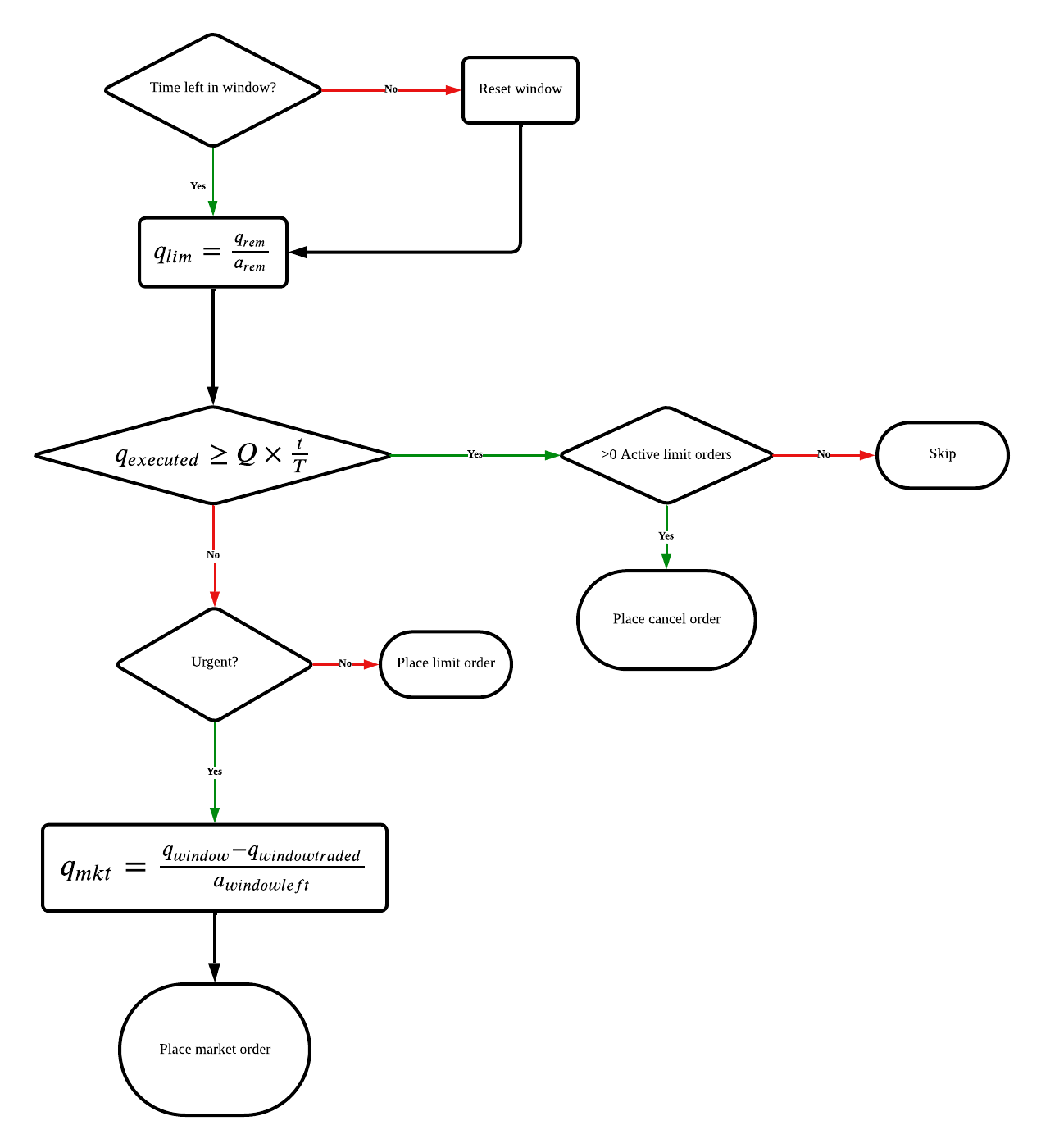}
    \caption{TWAP Agent Flowchart}
    \label{fig:twap_agent_flowchart}
\end{figure}

The algorithm that determines what action the TWAP agent takes is seen in Figure \ref{fig:twap_agent_flowchart}. The limit order quantity $q_{lim}$ is calculated with the ratio between the quantity remaining ($q_{rem} = Q - q_{executed}$), and the number of actions remaining for the entire execution time ($a_{rem}$). Actions are skipped and active LOs are cancelled if the agent has over-executed for the current time $t$. LOs are placed if the agent is not in the 'Urgent' state, and MOs are placed when the agent is in the 'Urgent' state---to ensure that the window quantity is successfully executed. The agent is in the `Urgent' state when 75\% of the trading window time has elapsed and the agent's LOs have execution rates so far of less than 90\%. 

\begin{figure}
    \centering
    \includegraphics[width=0.9\linewidth]{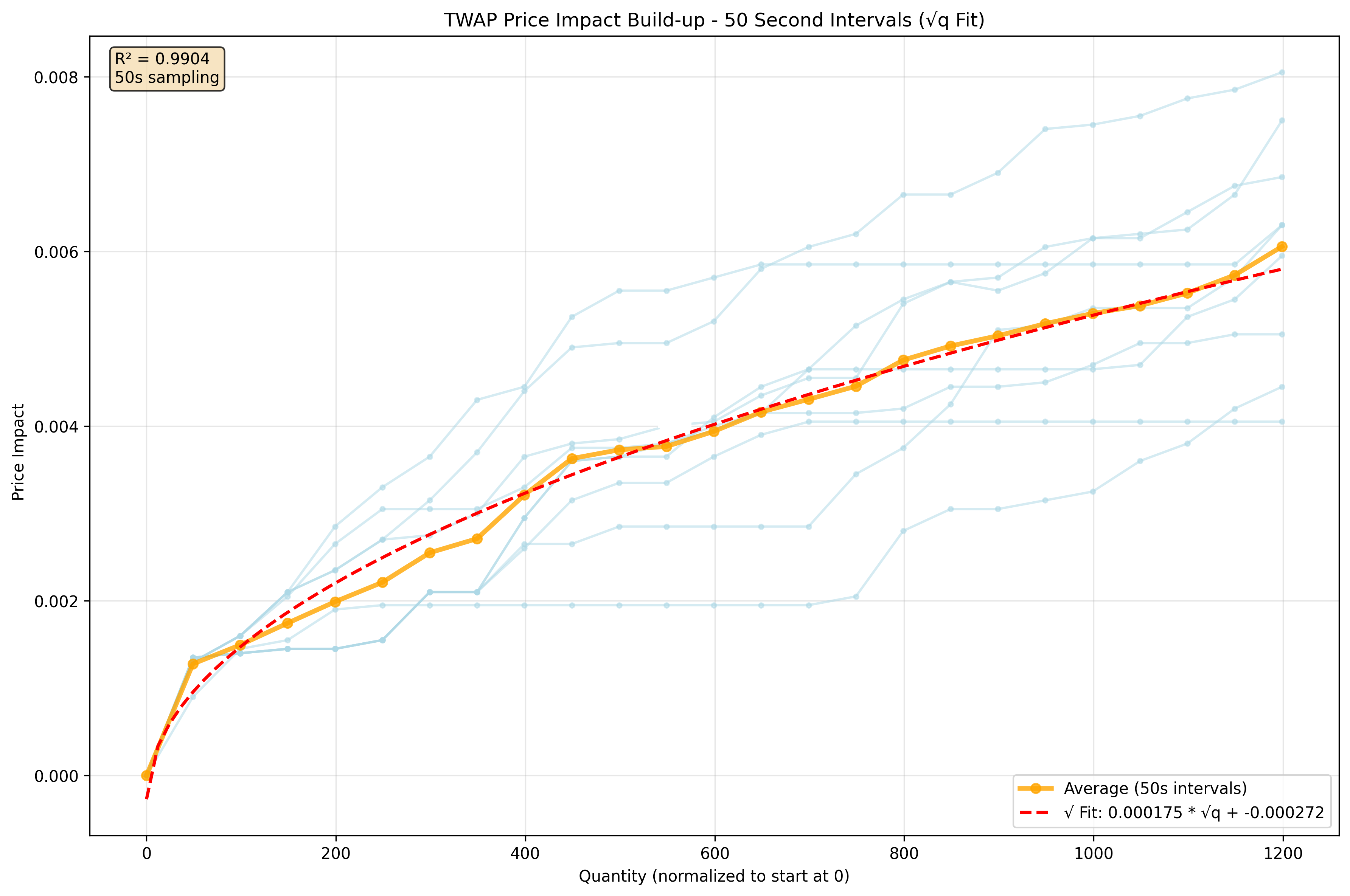}
    \caption{TWAP Agent Quantity vs Price impact During Execution}
    \label{fig:TWAP_predecay_50s_intervals}
\end{figure}

This implementation of TWAP in the limit order book simulation achieves the stylised facts of meta-orders' market impact as seen in \cite{maitrier2025generatingrealisticmetaorderspublic}.  Firstly, as shown in Figure \ref{fig:TWAP_predecay_50s_intervals}, the square root law (SQL) holds inside the metaorder. The relative price impact, as measured by $\frac{P_t - P_0}{P_0}$ is a square root function of the quantity executed so far, for $P_t$ the price at time step $t$ and $P_0$ the initial price. In the meta-order market impact testing, the TWAP agent began trading 100 seconds after the simulation started in order to let the market "warm up". The TWAP agent's execution period was 1200 seconds for a total quantity of 1200, and the window size was set to 50 seconds.

\par{Apart from the SQL, the other stylised fact of meta-orders' impact that was achieved was the concave price impact, or price 'relaxation' post metaorder execution. \cite{maitrier2025generatingrealisticmetaorderspublic} and \cite{bucci_et_al_2019} have shown that post execution price decay is concave. Simplistically, the post execution impact can be modeled by a decay, $G(t) \approx t^{-\beta}$ \cite{trades_quotes_bouchaud_book}. Additionally, empirical research has shown that for $z = \frac{t}{T}$, with $t=0$ being the start of execution and T defined as the total execution time (1200 in our tests), the impact can be represented by the function $\mathcal{I}_{prop} (Q, z) = z^{1-\beta} - (z-1)^{1-\beta}$ \cite{maitrier2025generatingrealisticmetaorderspublic}, \cite{bucci_et_al_2019}. }

\begin{figure}[H]
    \centering
    \includegraphics[width=0.9\linewidth]{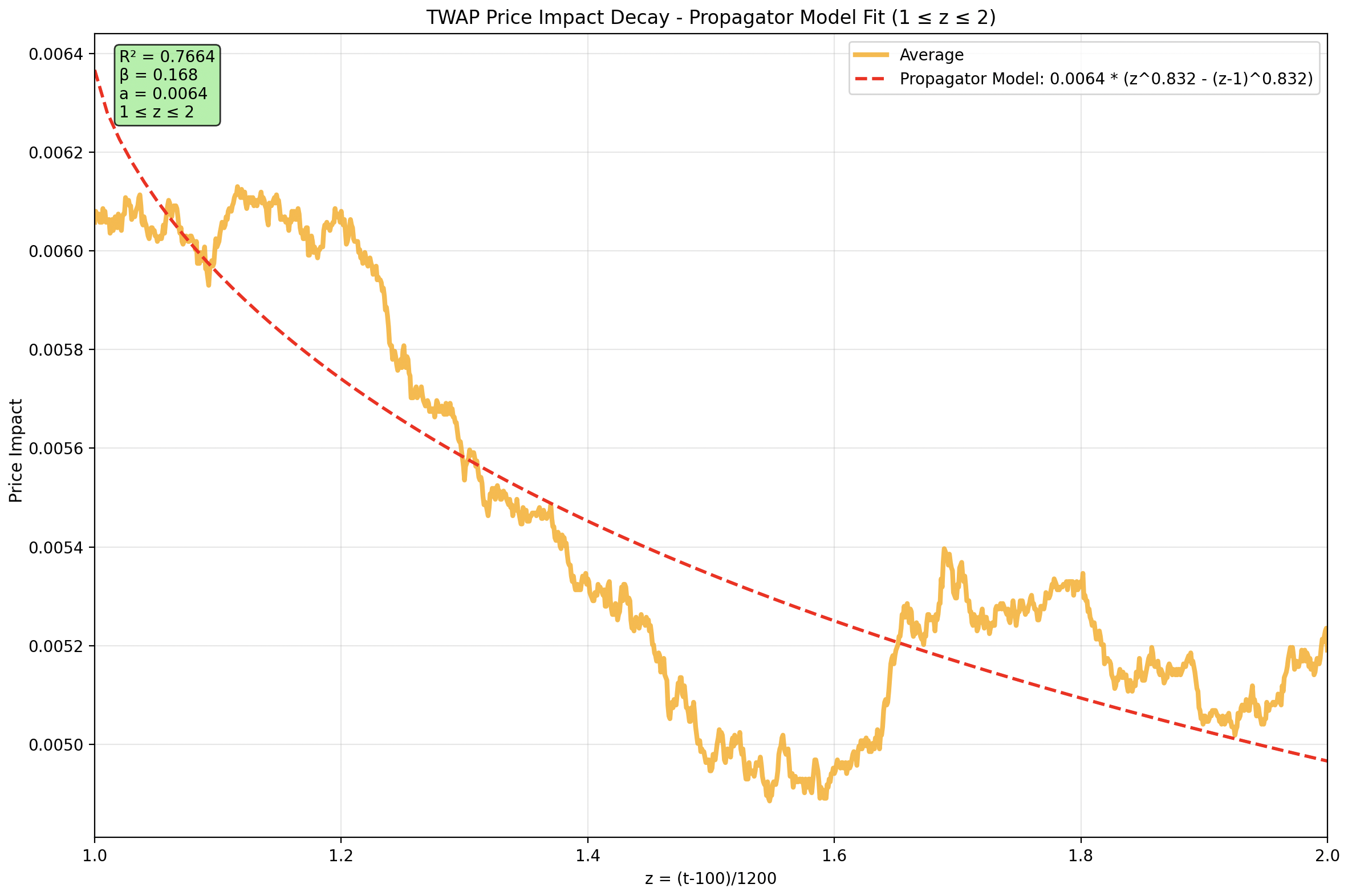}
    \caption{TWAP Price Impact Decay Post Execution}
    \label{fig:TWAP_postdecay}
\end{figure}

\par{As seen in Figure 3, the function is a good match for the post impact price decay for the TWAP agent. Additionally, our $\beta$ of 0.168 is very close to the 0.19 found by \cite{maitrier2025generatingrealisticmetaorderspublic}. }

\subsubsection{Testing TWAP Base Slippage}

In order to compare the performance of TWAP when trading alone versus when trading against the MM, we use slippage. Specifically, we calculate slippage using the target arrival method. When TWAP starts trading, the midprice is recorded (arrival price), and the final execution cost is compared with the cost if the entire order was executed at the arrival price. 

To test the slippage of the TWAP agent trading alone, two simulations were ran.

For the first, the total time of the simulation was 300 seconds of trading time with an additional 100 seconds of simulation warm up time. The agent was initialised to execute an order of size $Q = 300$ with action frequency $f = 1$, meaning an average child order size $q = 1$. The TWAP agent's window size was set to 50 seconds. In this experiment, the average market volume is $v = 0.55$ per second without trades from the TWAP agent. The additional market volume as a result of TWAP's behaviour can be calculated as TWAP's percentage of volume, or participation rate (POV), $100\frac{q}{v} \approx 181.81\%$. The POV rate being this significant explains the larger magnitude of the high POV (hPOV) slippage seen in Table \ref{tab:TWAP alone slippage}.

Thus, for the second TWAP base slippage experiment, the simulation parameters were changed so that the average market volume increased, decreasing the POV rate of TWAP to a regular amount. Specifically, the average market volume was increased by a factor of 40, becoming $v = 22$ for when the TWAP agent is not trading. Thus, the TWAP agent used had $q = 1, T = 300$, but with a much lower POV of $100\frac{q}{v} \approx 4.55$. This implementation of TWAP is henceforth referred to as rPOV$_1$.

\begin{table}[H]
    \centering
    \fontsize{9}{12}\selectfont
    \begin{tabular}{|c|c|c|c|}
        \hline
         &Buy Slippage & Sell Slippage & POV\\
         \hline
         hPOV & 18.68 & 13.92 & 180.81\%\\
         \hline
         rPOV$_1$ & 7.05 & 10.36 & 4.55\%\\
         \hline
    \end{tabular}
    \caption{rPOV$_1$ and hPOV TWAP Baseline Slippage}
    \label{tab:TWAP alone slippage}
\end{table}

The hPOV implementation of TWAP incurs a higher slippage when trading alone, which makes intuitive sense. 

\subsection{RL MM Agent}

As mentioned, the reinforcement learning model used for the RL MM trading agent is a model-free RL approximation for the HJB-QVI, trained using Proximal Policy Optimisation (PPO). PPO is a policy gradient method able to approximate in continuous time, and one that is robust to noisy updates \cite{jain2025impulse} \cite{SchulmanPPO}. 

To allow the RL agent to both optimise the action it takes when called but also the decision as to whether an action should be taken, two networks are used for the RL model \cite{jain2025impulse}. The decision network $d_\theta$ decides when the RL model should make an action, and the action network $u_\phi$ decides which action to take. The actions were restricted to top-of-book limit order placements and cancellations in order to simplify the RL agent's decision making. To prevent "costless unrealistic round-trips", we add a transaction fee feature to the simulation, charging 1 basis point (bps) once inventory is liquidated \cite{jain2025impulse}. 

To mitigate the variance that PPO training generally incurs, self-imitation learning (SIL) was also integrated into training for the RL model \cite{OhGuoSelfImitation}. In essence, it will encourage the RL to mimic previous trajectories that obtained a higher value than the value function's estimation for the current trajectory \cite{jain2025impulse}. 

The RL MM agent only has access to information regarding its own inventory, and the state of the LOB simulation \cite{jain2025impulse}. As the RL model has no extra knowledge of the TWAP agent's presence, this RL agent is referred to as the unaware RL agent (uRL). 

\subsubsection{uRL Agent Performance}
When trading by itself for 300 seconds with $f = 0.213$, after 100 seconds of simulation warm-up time, the uRL agent performs well, as measured by Sharpe ratio. 
Under these market conditions and agent initialisations, the base model from \cite{jain2025impulse}, without any additional training achieved an extremely high mean Sharpe of 129.089111 over 10 episodes. 

\section{uRL Agent vs TWAP Agent}
When testing the uRL agent against the regular and high POV (hPOV) TWAP agents in the Hawkes LOB simulation, the results in Table \ref{uRL performance table} were obtained for high and regular POV buy (B) and sell (S) agents. 

When testing against uRL, the regular POV TWAP agent was implemented in a different way. Instead of changing the simulation parameters as in the implementation of rPOV$_{1}$, which would involve changing it from the environment uRL was trained in, the TWAP parameters were changed. Action frequency, total quantity, and total time were scaled to achieve a POV rate of $\approx5\%$. Specifically, they were scaled to $f = 40$, $Q = 8$ and $T = 320$, with a window size of 160 seconds and, again, 100 seconds of simulation warm-up time. This implementation of TWAP is henceforth referred to as rPOV$_2$.

\begin{table}[H]
    {
    \fontsize{9}{12}\selectfont
    \centering
    \begin{tabular}{|c|c|c|c|c|}
        \hline
         &rPOV$_2$ B &rPOV$_2$ S & hPOV B & hPOV S \\
         \hline
         Sharpe&58.74 & 4.44 & -95.89 & -92.33\\
         \hline
    \end{tabular}
    \caption{uRL Annual Sharpe Against Different TWAP Agents}
    \label{uRL performance table}
    }
\end{table}

As can be seen in Table \ref{uRL performance table}, the uRL agent struggles to perform against the hPOV TWAP agent. This is intuitive, as the hPOV TWAP agent acts with a much higher frequency and with much more price impact than rPOV$_2$. 

It is worth noting, however, that despite its relatively low overall impact and activity, the rPOV$_2$ TWAP agent's presence does still affect the Sharpe of the uRL agent. The sell meta-order particularly impacts uRL's performance when compared to its Sharpe when trading alone. 

As it is clear the uRL agent is unable to trade effectively against the hPOV TWAP agent, slippage testing was only conducted against the rPOV$_2$ TWAP agent. Slippage for rPOV$_2$ TWAP trading against uRL was measured over the same episodes of testing, resulting in the data in Table \ref{uRLvsTWAP slippage table}.

\begin{table}[H]
    {
    \fontsize{9}{12}\selectfont
    \centering
    \begin{tabular}{|c|c|c|}
        \hline
         &Buy & Sell\\
         \hline
         Slippage&-1.04&1.63\\
         \hline
    \end{tabular}
    \caption{rPOV$_2$ slippage when trading against uRL}
    \label{uRLvsTWAP slippage table}
    }
\end{table}

As seen, the uRL MM keeps slippages relatively low for this rPOV$_2$ implementation of the regular POV TWAP agent. 

That being said, with the action frequency for uRL being $f = 0.213$, the very slow action frequency of the rPOV$_2$ TWAP agent of $f = 40$ does not mimic the difference in frequencies between MFT and HFT firms in the real world. On the other hand, the difference between the action frequency of uRL and the hPOV TWAP agent's action frequency of $f = 1$ does mimic real-world trade frequency differences between a HFT market maker and a MFT meta-order execution agent.

\section{Training RL Agent}

Because the action frequency of the rPOV$_{2}$ implementation of TWAP makes it a weaker representation of a meta-order execution agent, and because its low execution volume means it is possible the RL agent simply learns to trade around the TWAP's trades, for training, the implementation of TWAP used was rPOV$_{1}$. The baseline, then, for TWAP slippage comparison is the buy and sell slippage seen in Table \ref{tab:TWAP alone slippage}. Other than the TWAP agent's presence, the RL training setup was largely the same as the training for uRL. 

To train the RL agent, additional information was added to its state space: the TWAPPresent variable $\varrho$, defined in Table \ref{TWAPPresent def}. Effectively, this gives the RL agent full knowledge of the TWAP agent's behaviours, and thus we define the RL trained in this way as fRL i.e. 'fully aware' RL. 

\begin{table}[H]
    {
    \fontsize{9}{12}\selectfont
    \centering
    \begin{tabular}{|c|c|c|c|}
        \hline
         State & No TWAP & TWAP Buy & TWAP Sell \\
         \hline
         $\varrho$& 0  & 1 & -1\\
         \hline
    \end{tabular}
    \caption{States for $\varrho$}
    \label{TWAPPresent def}
    }
\end{table}

This additional information also essentially gives the fRL agent knowledge as to the market's price trajectory. It would also follow that the fRL agent would front-load its inventory for the buy signal, and reduce its inventory for the sell signal, front-running the TWAP meta order execution agent. Yet, this does not exactly describe the results we obtained from training.

To ensure that the fRL agent both learns how to trade against TWAP and learns how to trade in regular market conditions, training was conducted under several structures. 

The first was to randomise the TWAP starting time using a normal distribution with $\mu = 150$, for 300 seconds of trading time. Training was then changed to have TWAP always starting at 150 seconds, halfway through the RL agent's trading time. To maintain $q=1$, the RL agent is still initialised as if it had $T = 300$, regardless of its actual starting time. 
\begin{figure}
    \centering
    \includegraphics[width=0.95\linewidth]{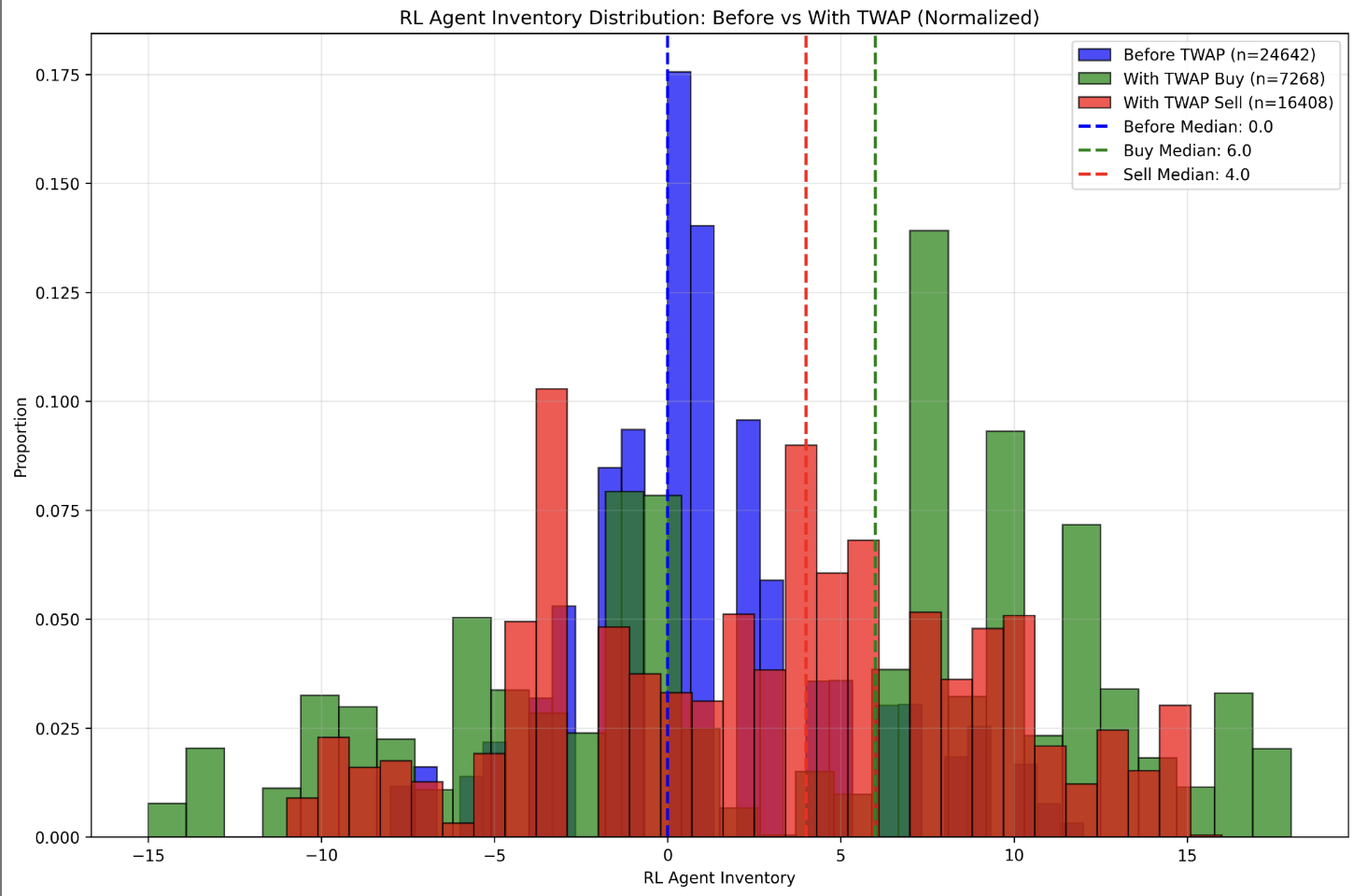}
    \caption{fRL Agent Inventory Distribution, 140 episodes}
    \label{fig:140epochsInventorydist}
\end{figure}
Figure \ref{fig:140epochsInventorydist} shows the fRL agent's inventory distribution after 140 episodes of training. It is observed that the median inventory for the fRL agent with $\varrho = 1$ is 6, and the median inventory with $\varrho = -1$ is 4. Both are above the fRL agent's median inventory of 0 with $\varrho = 0$. This means that it is behaving as expected for buy episodes but is not for sell episodes, implying that its performance may not be as high for buy episodes than for sell episodes. 

\begin{figure}[H]
    \centering
    \includegraphics[width=1\linewidth]{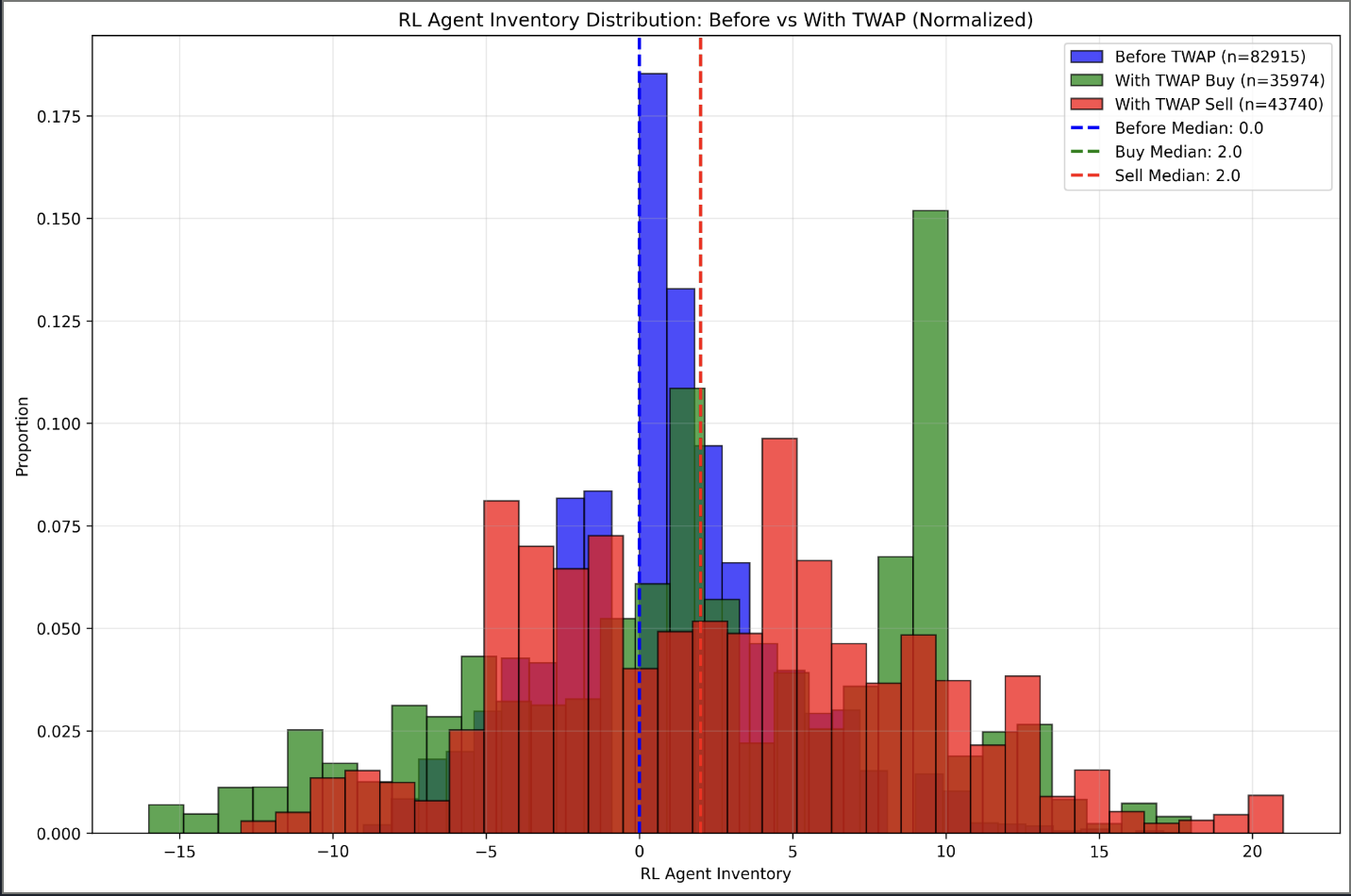}
    \caption{fRL Agent Inventory Distribution, 246 episodes}
    \label{fig:highepochsInventorydist}
\end{figure}

As seen in Figure \ref{fig:highepochsInventorydist}, with additional training, both the buy and sell median inventories decrease and converge to 2. The fRL model, after additional episodes of training, did not approach simulating the expected behaviour of a market maker. Thus, the comprehensive performance testing against the TWAP agents that follows is conducted with the fRL agent after 140 episodes of training only. 

\subsection{Results: fRL Agent vs TWAP Agent}
The fRL agent was also tested against buy and sell TWAP agents, in the same manner of the majority of its training episodes. That is, the TWAP agent starts trading halfway through fRL's trading time, at 150 seconds. 

\subsubsection{fRL Agent Performance}
The fRL's performance was measured over 20 episodes of trading against the buy TWAP agent, and 20 episodes of trading against the sell TWAP agent. We again use Sharpe ratio to measure the performance of the fRL trading agent.

\begin{table}[H]
    \centering
    \fontsize{9}{12}\selectfont
    \begin{tabular}{|c|c|c|}
        \hline
         & Buy & Sell \\
        \hline
        Before TWAP & 39.63 & 38.18 \\
        \hline
        During TWAP & 150.43 & -4.96 \\
        \hline
    \end{tabular}
    \caption{fRL Sharpe before and during TWAP}
    \label{tab:fRL performance vs TWAP}
\end{table}

Table \ref{tab:fRL performance vs TWAP} presents the mean Sharpe ratio across all test episodes. The “before” period corresponds to the time window when only the fRL agent is active in the market, while the “during” period captures the interval in which both the fRL and TWAP agents are executing simultaneously. The results reveal a notable asymmetry in the fRL agent’s performance when facing the different TWAP agent sides. 

As seen, fRL's performance increases significantly during the TWAP buy phases, showing it has learnt to capitalise on the buy meta-order execution and the resulting upward pressure on price that the TWAP would cause. However, this same improvement was not seen for the sell episodes. The performance of fRL is significantly higher for buy episodes when $\varrho = 1$ than for sell episodes when $\varrho = -1$.

Indeed, the performance of fRL decreases during sell episodes as compared to before them, showing that the agent had not yet learned to capitalise on the sell-side meta-order flows. This discrepancy between the performances of fRL against the two sides of TWAP matches the intuition from the inventory data of fRL observed during training: that the fRL agent's bias toward long positions and its excessive inventory mean it performs worse when trading against sell-side TWAP. Because testing uRL in the updated simulator environment with higher average market volumes would not be possible due to the uRL agent being trained for smaller average market volumes, a direct comparison is difficult. However, when considering the improvement of the fRL agent with the training against TWAP, we can use the uRL's performance against the rPOV$_2$ and hPOV TWAP agents as seen in Table \ref{uRL performance table} as loose benchmarks. From this, we can see that the agent has improved in its performance when trading against TWAP significantly, for both buy and sell TWAP agents. 

In sum, the fRL agent has learnt asymmetrically: it thrives when the market is characterised by upward price momentum driven by the buy TWAP agent but struggles to profit under sustained sell-side pressure. This highlights a key area for potential improvement for the fRL.

\subsubsection{TWAP Agent Slippage}

The TWAP agent’s slippage, measured in bps, was also evaluated during the same testing episodes to assess the impact of the fRL MM agent’s activity on the cost of execution for the metaorder.

\begin{table}[H]
    \centering
    \fontsize{9}{12}\selectfont
    \begin{tabular}{|c|c|}
        \hline
         Buy Slippage & Sell Slippage \\
         \hline
         0.24 & 2.31 \\
         \hline
    \end{tabular}
    \caption{TWAP Slippage vs fRL, in bps}
    \label{tab:TWAP fRL slippage}
\end{table}

As seen in Table \ref{tab:TWAP fRL slippage}, though the MM has clearly learnt to profit from trading against the buy side TWAP agent, the results from the same testing episodes showed that the TWAP agent's slippage did not increase significantly. In fact, when comparing with Table \ref{tab:TWAP alone slippage}, it appears that trading against fRL decreased execution costs for TWAP. The slippage decreased, from 7.05 to 0.24. This means that the fRL MM may have learned strategies that stabilised market prices.

Additionally, while the sell TWAP slippage of 2.31 is lower than the sell TWAP slippage when the agent was trading alone of 10.36, the sell TWAP slippage is higher than the buy TWAP slippage. This, combined with the worse performance of fRL when trading against sell TWAP when compared to fRL's performance when trading against buy TWAP, may imply a sort of inverse relationship between TWAP slippages and fRL profit when trading against the slippages. 

This is interesting, because it prompts the question: if the fRL agent's significantly increased profit does not cause increased costs for buy metaorders executed using TWAP, what strategy is the fRL agent employing, and, is it more profitable than front-running the meta-order? 

\section{Discussion and Conclusion}

In this paper, we investigated how reinforcement learning-based market making agents interact with medium-frequency traders executing metaorders within a Hawkes limit order book environment. Using the compound Hawkes process to model the endogenous nature of order flow allowed us to effectively replicate the dynamics of adverse selection between high-frequency traders and medium-frequency participants observed empirically. The impulse control reinforcement learning framework, implemented through PPO and self-imitation learning, enabled the agent to approximate the solution to the high-dimensional HJB-QVI and to learn intervention decisions analogous to those of real-world high-frequency market makers.

The results show that the unaware RL (uRL) agent performs well in isolation but exhibits varying sensitivity when trading against the TWAP meta-order execution agent. Against high participation rate (hPOV) TWAP agents, the uRL’s Sharpe ratio decreased significantly, consistent with the greater price impact and frequency mismatch. Introducing state information about the TWAP’s presence and direction led to the fully informed RL (fRL) agent. During training, the fRL agent learned to exploit buy-side metaorder executions effectively, as evidenced by the large increase in Sharpe ratio during TWAP buy phases. However, the performance improvement was not symmetric across buy and sell episodes, with inventory distributions indicating stronger adaptation to buy phases than to sell phases.

Interestingly, despite the fRL agent’s higher profitability during TWAP buy executions, the slippage experienced by the TWAP agent did not increase significantly. This suggests that while the RL market maker learned to capitalise on predictable order flow, the profits did not necessarily stem from worsening execution costs for the medium-frequency trader. The fRL agent’s strategy may therefore involve more efficient liquidity provision or improved inventory management rather than direct adverse selection.

Overall, these findings provide an initial framework for understanding how reinforcement learning-driven trading agents behave in adversarial multi-agent market environments with endogenous dynamics. As high-frequency trading continues to proliferate, further research into defensive mechanisms for medium-frequency traders and extensions to more sophisticated metaorder execution strategies will be crucial to mitigating potential adverse selection in increasingly AI-dominated markets.

\section{Future Works}

The interesting performance and slippage results of this paper highlight a number of avenues for future investigation. Firstly, to investigate the strategy employed by fRL further and determine if the profitability of maintaining steady prices is higher than front-running the meta-order execution agent. Further training for fRL against the sell side TWAP agent is needed, but it would also be informative to investigate how the MM RL learns when training against a different meta-order execution strategy, such as percentage of volume (POV) or volume-weighted average price (VWAP). POV and VWAP are slightly more sophisticated execution strategies, aiming to execute at a specific percentage of market volume (real market volume in the case of POV, historical/estimated market volume in the case of VWAP) and theoretically minimising market impact \cite{algotradingquantstrats}. 

In order to improve the RL MM's performance, a slightly different implementation of the RL agent could be used. If the state space for the RL agent included the current time, the RL agent would be able to predict when the TWAP agent would place an order next, increasing the likelihood it would learn a strategy that is adversarial. In order to make this work effectively, the actor-critic multi-layer perception (MLP) neural network would be replaced with a long short-term memory (LSTM) neural network. This is because MLP is poor at capturing temporal dependencies, while LSTM maintains memory of previous actions, making better use of the added current time information. 

If additional and reworked training of the RL MM agent results in adverse selection causing increased slippages for the meta-order execution agent, defensive mechanisms for the meta-order execution agent will be developed and investigated. Recent work has shown the application of recurrent neural networks (RNN) to execution \cite{genet2025lemsprimerlargeexecution}, and further investigation into the ability for large execution models (LEMs) to reduce execution slippage when trading against an adversarial agent is needed. More deterministic defensive strategies may also be effective, such as a slippage detection function in the meta-order execution implementation which halts execution temporarily or changes strategy in order to confuse the RL MM agent. 

Finally, research into developing the RL MM agent's ability to identify the presence of meta-orders in the market may be useful for the investigation into how meta-order executions are identified for adverse selection.

\section{Acknowledgments}
Konark Jain would like to acknowledge JP Morgan Chase \& Co. for his PhD scholarship. Finally,  we  are  grateful  to  the anonymous reviewers for their constructive feedback.
{\small \section*{Disclaimer}
Opinions and estimates constitute our judgement as of the date of this Material, are for informational purposes only and are subject to change without notice. This Material is not the product of J.P. Morgan’s Research Department and therefore, has not been prepared in accordance with legal requirements to promote the independence of research, including but not limited to, the prohibition on the dealing ahead of the dissemination of investment research. This Material is not intended as research, a recommendation, advice, offer or solicitation for the purchase or sale of any financial product or service, or to be used in any way for evaluating the merits of participating in any transaction. It is not a research report and is not intended as such. Past performance is not indicative of future results. Please consult your own advisors regarding legal, tax, accounting or any other aspects including suitability implications for your particular circumstances. J.P. Morgan disclaims any responsibility or liability whatsoever for the quality, accuracy or completeness of the information herein, and for any reliance on, or use of this material in any way.
Important disclosures at: www.jpmorgan.com/disclosures }
\bibliography{aaai2026}

\end{document}